\documentclass{aa}
\usepackage{graphicx}

\begin{document}

  \title{Neural networks and photometric redshifts}

   \author{R. Tagliaferri
          \inst{1} , \inst{2}
          \and
          G. Longo \inst{3}
          \and 
          S. Andreon \inst{4}
          \and 
          S. Capozziello \inst{5}
          \and 
          C. Donalek \inst{4}
          \and 
          G. Giordano \inst{3}
          }
   \offprints{G. Longo}

   \institute{DMI - University of Salerno, 84081, Baronissi (SA), Italy\\
         \and
             INFM, Unit\`{a} di Salerno, 84081 Baronissi, Italy\\
         \and 
             Department of Physical Sciences, University Federico II\ of Naples, I-80126, Italy\\
	   \and           
             Osservatorio Astronomico di Capodimonte, via Moiariello 16, 80131, Napoli\\
         \and 
             Dipartimento di Fisica, Universit\`{a} di Salerno, Baronissi, Italy\\
             \thanks{}
             }

   \date{Received xxxxxx, 2002; accepted xxxxx}

\abstract{
We present a neural network based approach to the determination of
photometric redshift. The method was tested on the Sloan Digital Sky Survey
Early Data Release (SDSS-EDR) reaching an accuracy comparable and, in some
cases, better than SED template fitting techniques. Different neural
networks architecture have been tested and the combination of a Multi\ Layer
Perceptron with 1 hidden layer (22 neurons)\ operated in a Bayesian
framework, with a Self Organizing Map used to estimate the accuracy of the
results, turned out to be the most effective. In the best experiment, the
implemented network reached an accuracy of 0.020 (interquartile error) in
the range $0<z_{phot}<0.3$, and of 0.022 in the range $0<z_{phot}<0.5$.
}

   \maketitle

\section{Introduction}

\noindent

Ongoing and planned digital surveys such as the Sloan Digital Sky Survey and
the implementation of an International Virtual Observatory, will enormously
increase the quality and the amount of data available to the astronomical
community. The scientific exploitation of these data, while promising new
answers to old questions, is also stimulating the implementation of new
tools capable to deal in an effective way with unprecedently large volumes
of data. One field which will be deeply affected by these new data sets is
that of the large scale of the universe. Planned or ongoing surveys such as
the Sloan Digital Sky Survey (SDSS,;York et al. 2000), the VIRMOS-VLT\
Survey (Le F\`{e}vre et al. 2000),\ the VST\ Survey (G. Busarello, private
communication) will provide a huge amount of high accuracy spectroscopic and
photometric data which will enable observational cosmologists to map with
unprecedented accuracy and detail the properties and the structure of the
Universe.

In this respect it needs to be stressed that, in spite of the recent and
well known advances in multiobject spectroscopy, photometric redshifts (cf.
Baum 1962, Pushell et al. 1982)\ derived from multicolor photometry have
been for a long time and still are the only tool which may be effectively
used to evaluate the distances of large number of galaxies. Most photometric
redshifts methods rely on a $\chi ^{2}$ fitting of a library of template
Spectral Energy Distributions (hereafter SEDs) to the observed data points,
and differ mainly in how the SEDs are derived and on how they are fitted to
the data.

SEDs may either be derived from population synthesis models (cf. Bruzual and
Charlot 1993) or be spectra of real objects selected in order to ensure a
sufficient coverage of morphological types and/or luminosity classes. As
effectively stressed by Koo (1999), both approaches (synthetic and
empirical) have their pro's and con's.

Synthetic spectra, for instance, sample an 'a priori' defined grid of
mixtures of stellar populations and may either include unrealistic
combinations of parameters or exclude some unknown cases, while empirical
templates are usually derived from nearby and bright galaxies and may
therefore be not representative of the spectral properties of galaxies
falling in other redshift ranges. The various methods are extensively
compared and discussed in several papers (cf. Koo 1999;\ Fernandez-Soto et
al. 2001) and have been applied to many different data sets such as the
Hubble Deep Field (cf. Massarotti et al. 2001a, 2001b) .

Another approach, which is in the same line of the one discussed in this
paper, can be applied only to what we shall call 'mixed surveys', \textit{id
est} datasets where accurate and multiband photometric data for a large
number of objects are supplemented by spectroscopic redshifts for a small
but statistically significant subsample of the same objects. In this case,
the spectroscopic data can be used to constrain the fit of a polynomial
function mapping the photometric data (cf. Connolly et al. 1995, Wang et al.
1998, Brunner et al. 2000). It needs to be stressed that, at difference with
the SED fitting methods, this interpolative approach cannot be effectively
applied to objects fainter than the spectroscopic limit since, in absence of
an 'a priori' knowledge, impossible extrapolations would be required. It
could be argued that the needed 'a priori' knowledge could be extracted from
population synthesis models, but it is apparent that, in this case, the
uncertainties of the two methods would add up and SED's fitting methods
would - in any case - be more accurate and preferable.

Interpolative methods, however, offer the great advantage that they are
trained on the real\ Universe and do not require strong assumptions on the
physics of the formation and evolution of stellar populations. Neural
Networks (hereafter NNs)\ are known to be excellent tools for interpolating
data and for extracting patterns and trends (cf. the standard textbook by
Bishop 1995) and in this paper, we shall discuss the application of a set of
neural tools to the determination of photometric redshifts in large ''mixed
surveys'' (cf. Giordano 2001). In section 2 we introduce the basic concepts
of Neural Networks paying special attention to the Multi Layer Perceptron
and the Self Organising Maps;\ in Section 3 we discuss a first application
to the SDSS Early Data Release data (Stoughton et al. 2001) and in Section 4
we show how SOM\ can be used to evaluate the degree of contamination of the
final redshift catalogues. In Section 5, finally we shall draw our
conclusions and discuss some possible applications of the neural tools. In a
forthcoming paper we shall compare the results of different photometric
redshifts methods applied to the same SDSS-EDR data and will discuss the
properties of the photometric redshift catalogue derived with the method
described in this paper (Longo et al. 2002, in preparation).

\section{Neural Networks}

NNs, over the years, have proven to be a very powerful tool capable to
extract reliable information and patterns from large amounts of data even in
the absence of models describing the data (cf. Bishop 1995) and are finding
a wide range of applications also in the astronomical community:\ catalogue
extraction (Andreon et al. 2001), star/galaxy classification (Bertin and
Arnout, 1996, Andreon et al. 2001), galaxy morphology (Storrie-Lombardi et
al. 1992;\ Lahav et al. 1996), classification of stellar spectra (Bailer and
Jones 1998, Allende Prieto et al. 2000, Weaver 2000), data quality and data
mining (Tagliaferri et al. 2002).

A NN is usually structured into an input layer of neurons, one or more
''hidden'' layers and one output layer. Neurons belonging to adiacent layers
are usually fully connected and the various types and architectures\ of the
NNs are identified by the different topologies adopted for the connections
and by the choice of the activation function (details can be found in the
standard book by Bishop 1995). From the operational point of view, NNs can
be divided into two main types, supervised and unsupervised systems,
accordingly to the type of learning.

In the first case, NNs learn how to recognise a specific pattern or
characteristic on a set (hereafter ''training set'')\ of labeled data
containing for each input vector also the desired output (hereafter
''target''). In unsupervised systems, instead, NNs look for the statistical
similarity of the imput data. In this work both types of NNs are used to
perform different tasks.

The AstroMining software (Longo et al. 2001)\ is a package written in the
MatLab \copyright\  environment to perform a large number of data mining and
knowledge discovery tasks, both supervised and unsupervised, in large
multiparametric astronomical datasets. The package relies also on the
Matlab \copyright\  ''Neural Network'', the ''SOM'' (Vesanto 1997)\ and the
''Netlab'' (Nabney and Bishop 1998) toolboxes.

AstroMining accepts as input any ASCII table containing a header describing
the contents of each colum and then a set of parameters. Via interactive
interfaces, it is possible to perform a large number of operations:\ i)\
manipulation of the input data sets;\ ii)\ selection of relevant
parameters;\ iii)\ selection of the type of neural architecture;\ iv)
selection of the training validation and test sets construction procedure;\
v) etc. The package is completed by a large set of visualization and
statistical tools which allow to estimate the reliability of the results and
the performances of the network. The user friendliness and the generality of
the package allow both a wide range of applications and the easy execution
of experiments (more details on other aspects of the AstroMining tool which
are not relevant to the present work may be found in Tagliaferri et al.
2002).

Let us focus now on some fundamental aspects connected with the use of
supervised NNs. In order to perform correctly, almost all supervised NNs
need to be trained, validated and tested on three independent datasets. In
order to achieve good generalization performances, the training set needs to
be representative of the typical data which will be passed to the network in
the application phase. The validation data set (which is often and
erroneously ignored in many NN applications)\ is a second dataset disjoined
from the training set but having the same statistical properties. The role
played by the validation set is subtle but crucial:\ by using the training
set alone, in fact, the NNs (and the MLP\ in particular)\ may easily run
into overfitting errors, thus loosing all generalization properties. It
needs to be stressed in fact, that while the error computed on the training
set may decrease asintotically, the capability of the network to reproduce
patterns not encountered during the training phase may decrease (this is
known as 'over-fitting' condition), in other words, the NN\ will learn how
to reproduce the patterns in the training set but will produce completely
wrong results when applied to other data sets. The validation set prevents
this from happening via a so called 'regularization technique': during the
training phase, at regular intervals, the training is interrupted (and the
weights of the neurons are frozen), then the net is run on the validation
set in order to compute the error with respect to the desired output; the
training is stopped when the error computed on the validation set shows a
significant increasing trend and, finally, the NN corresponding to the
minimum error is selected.

After this phase the final performances of the resulting weight
configuration are tested on a third data set, the so called 'test set',
which is, once more, completely disjoined from the previous two. Another
regularization method not requiring a validation set, namely the Bayesian
Learning Approach, will be described in the next paragraph.

\subsection{The Multi Layer Perceptron - MLP}

Due to its interpolation and capabilities, the Multi Layer Perceptron (MLP)\
is one of the most widely used neural architectures. As most other networks,
the MLP\ is structured into an input layer, one or more hidden layers and
one output layer (see figure 1). We implemented an MLP\ with one hidden
layer and $n$ imput neurons, where $n$ is the number of parameters selected
by the user as input in each experiment.

\begin{figure}
   \centering
   \includegraphics[width=10cm]{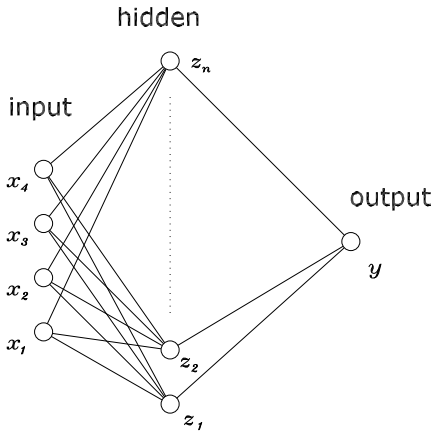}
     \caption{A schematic representation of the Multi Layer Perceptron}
     %\label{FigGam}%
\end{figure}

As just mentioned, it is also possible to train NN's in a Bayesian
framework, which allows to find the more efficient among a population of
NN's differing in the hyperparameters controlling the learning of the
network (cf. Bishop 1995), in the number of hidden nodes, etc. The most
important hyperparameters being the so called $\alpha $ and $\beta $. The
parameter $\alpha $ is related to the weights of the network: a larger value
for a component of $\alpha $ implying a less meaningful corresponding
weight, thus allowing to estimate the relative importance of the different
inputs\ (\textit{Automatic Relevance Determination};\ Bishop 1995) and,
therefore, the selection of the input parameters which are more relevant to
a given task. The parameter $\beta $ is instead related to the variance of
the noise: a smaller value corresponding to a larger value of the noise and
therefore to a lower reliability of the network. The Bayesian method allows
the values of the regularization coefficients to be selected using only the
training set, without the need for a validation set.

The implementation of a Bayesian framework requires several steps:\
initialization of weights and hyperparameters; training the network via a
non linear optimization algorithm in order to minimize the total error
function. Every few cycles of the algorithm, the hyperparameters are
re-estimated and eventually the cycles are reiterated.

\subsection{The Self Organizing Maps - SOM}

The Self-Organizing Map (SOM), developed by Kohonen (1995), is one of the
most used NN model. The SOM algorithm is based on unsupervised competitive
learning, \textit{id est} the training is entirely data-driven and all
neurons of the map compete with each other producing only one winning neuron
for each input vector. This property turns SOM into an ideal tool for KDD\
and expecially for its exploratory phase: data mining (Vesanto 1997). Among
various other advantages, SOM\ allow an approximation of the probability
density function of \ the training data, the derivation of prototype vectors
best describing the data, and a highly visualized and user friendly approach
to the investigation of the data.

A SOM is composed by neurons located on a regular, usually 1 or
2-dimensional grid. Each neuron $i$ of the SOM is represented by an $n$%
-dimensional weight or reference

$m_{i}=\left[ m_{i_{1}},m_{i_{2}},...,m_{i_{n}}\right] ^{T}$

where $n$ is the dimension of the input vectors. Higher dimensional grids
are not generally used since their visualization is much more problematic.
Usually the map topology is a rectangle \ but also toroidal topologies have
been used successfully. The neurons of the map are connected to adjacent
neurons by a neighborhood relation dictating the structure of the map. In
the 2-dimensional case the neurons of the map can be arranged either on a
rectangular or on a hexagonal lattice. The number of neurons determines the
granularity of the resulting mapping, which affects the accuracy and the
generalization capability of the SOM.

The use of SOM for data mining requires several steps : construction,
normalization and initialization of the Data Set, (unsupervised) training,
visualization of the resulting map, and, finally, analysis of the results.
The first two steps depend on the individual data set to be processed and
the normalization is made in order to achieve $mean=0$ and $variance=1$.

In the basic SOM algorithm, the topological relations and the number of
neurons are fixed from the beginning. The number of neurons should usually
be selected by trial and error, with the neighborhood size controlling the
smoothness and generalization of the mapping. Before training, in the course
of the initialization phase, initial values are given to the weight vectors.
The SOM is robust regarding the initialization, but if properly accomplished
it allows the algorithm to converge faster to a good solution. Typically,
any of the following initialization procedures may be used:

\begin{itemize}
\item  random initialization, with the weight vectors initialized with small
random values;

\item  sample initialization, where the weight vectors are initialized with
random samples drawn from the input data set;

\item  linear initialization, where the weight vectors are initialized in an
orderly fashion along the linear subspace spanned by the two principal
eigenvectors of the input data set. These eigenvectors can be calculated
using Gram-Schmidt procedure (Kohonen 1995).
\end{itemize}

\begin{figure*}
   \centering
   \includegraphics[width=14cm]{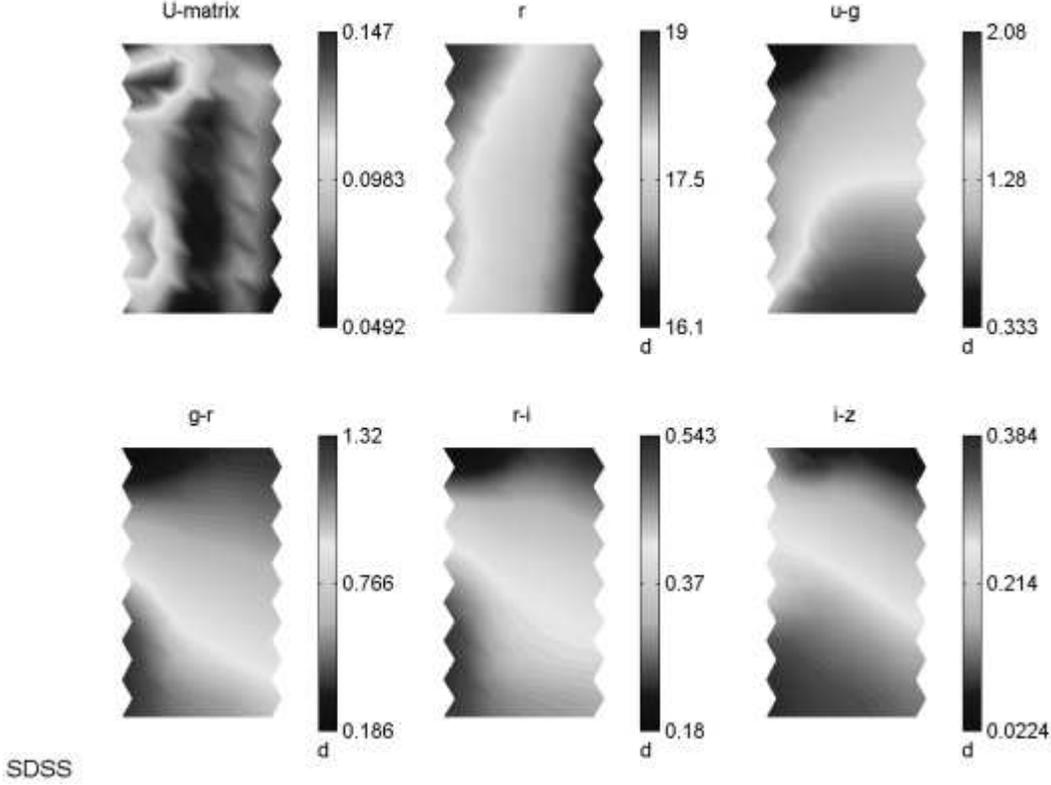}
     \caption{ U matrix and plane components of a 120 nodes SOM\ applied to the
SDSS-EDR\ data. The parameters are the r petrosian magnitude and the i-r,
g-r, r-i, i-z colors. Regions of uniform color are the clusters. In spite of
the rather uniform variation over the individual parameters, the U Matrix
clearly shows the existence of several clusters. For an interpretation of
the above, see Figure 5. }
     %\label{FigGam}%
\end{figure*}

During the training phase, one sample vector $\mathbf{x}$ from the input
data set is randomly chosen and a similarity measure is calculated between
it and all the weight vectors of the map. The Best-Matching Unit (BMU),
denoted as $c$, is the unit with weight vector having the greatest
similarity with the input sample $\mathbf{x}$. The similarity is usually
defined by means of a distance measure, typically an Euclidean distance.
Formally the BMU is defined as the neuron for which:

$\left\| \mathbf{x}-\mathbf{m}_{c}\right\| ={\min }_i \left\| 
\mathbf{x}-\mathbf{m}_{i}\right\| $

where $\left\| .\right\| $ denotes the distance measure. \ After finding the
BMU, the weight vectors of the SOM are updated. The weight vectors of the
BMU and its topological neighbors are moved in the direction of the input
vector, in the input space. The SOM update rule for the weight vector of the
unit $i$ is:

$\mathbf{m}_{i}\left( t+1\right) =\mathbf{m}_{i}\left( t\right)
+h_{ci}\left( t\right) \left[ \mathbf{x}\left( t\right) -\mathbf{m}%
_{i}\left( t\right) \right] $

Where $t$ denotes the time, $\mathbf{x}(t)$ is the input vector and $%
h_{ci}(t)$ denotes the neighborhood kernel around the winner unit. The
neighborhood kernel is a non-increasing function of time and of the distance
of unit $i$ from the winner unit $c$. It defines the region of influence
that the input sample has on the SOM. The kernel is composed by two parts:
the neighborhood function $h(d,t)$ and the learning rate function $\alpha
(t) $:

$h_{ci}\left( t\right) =h\left( \left\| \mathbf{r}_{c}-\mathbf{r}%
_{i}\right\| ,t\right) \alpha (t)$

where $r_{i}$ is the location of unit $i$ on the map grid. The neighborhood
function used in our experiments is the Gaussian neighborhood function:

$exp\left( \frac{-\left\| \mathbf{r}_{c}-\mathbf{r}_{i}\right\| ^{2}}{%
2\sigma ^{2}\left( t\right) }\right) $

The learning rate $\alpha (t)$ is a decreasing function of time. Two
commonly used forms are a linear function and a function inversely
proportional to time:

$\alpha (t)=\frac{A}{t+B}$

where $A$ and $B$ are some suitably selected constants. The training is
usually performed into two phases. In the first phase, relatively large
initial $\alpha $ value and neighborhood radius are used. In the second
phase both the $\alpha $ value and the neighborhood are small from the
beginning. This procedure corresponds to first tuning the SOM approximately
to the same space as the input data and then fine-tuning the map. The SOM\
toolbox (Vesanto 1997) includes the tools for the visualization and analysis
of SOM and, since the weight vectors are ordered on the grid, the
visualization of the U\ matrix turns out to be expecially useful in the data
understanding/survey phase. The U matrix visualizes the clustering
structures of the SOM as distances (in the assumed metric) between
neighboring map units, thus high values of the U-matrix indicate a cluster
border, uniform areas of low values indicate clusters themselves.

In Figure 2, we show both the U matrix for the whole data set, and the
structure of the individual components.

Another advantage of SOM is that it is relatively easy to label individual
data, \textit{id est} to identify which neuron is activated by a given input
vector. The utility of these properties of the SOM will become clear in the
next paragraphs.

\section{Application to the SDSS-EDR data}

A preliminary data release (Early Data Release or EDR) of the SDSS was made
available to the public in 2001 (Stoughton et al. 2001). This data sets
provides photometric, astrometric and morphological data for an estimated 16
millions of objects in two fields:\ an Equatorial $2^{circ}$ 
wide strip of constant declination centered around $\delta $=0 and a
rectangular patch overlapping with the SIRTF\ First Look Survey. 

\begin{figure}
   \centering
   \includegraphics[width=7cm]{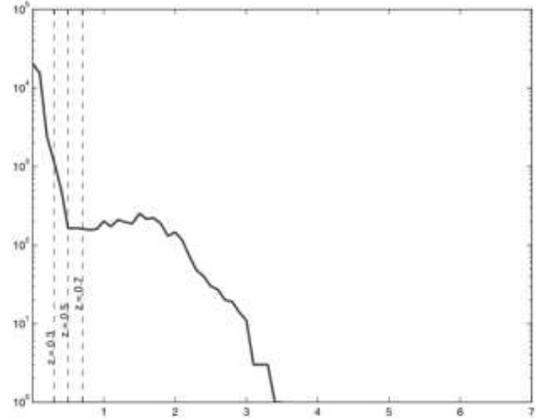}
     \caption{Logarithmic histogram of the redshifts in the SDSS-EDR. 
              Notice the sharp discontinuity at z=0.5.}
     %\label{FigGam}%
\end{figure}

The EDR provides also spectroscopic redshifts for a little more than $50.000$
galaxies distributed over a large redshift range and is therefore
representative of the type of the data which will be produced by the next
generation of large scale surveys. In order to build the training,
validation and test sets, we first extracted from the SDSS-EDR a set of
parameters ($u$, $g$, $r$, $i$, $z$, both total and petrosian magnitudes,
petrosian radii, $50\%$ and $90\%$\ petrosian flux levels, surface
brightness and extinction coefficients;\ Stoughton et al. 2001) for all
galaxies in the spectroscopic sample.

In this data set, redshifts are distributed in a very dishomogeneous way
over the range $0-7.0$ (Figure 3;\ $93\%$ of the objects have $z<0.7$).

\textbf{Table 1:}\ Cuts applied to magnitudes and colors.
\begin{tabular}{ll}
$15<u<21.5$ & $-1.5<u-g<5$ \\ 
$15.0<g<21.7$ & $-1<g-r<3$ \\ 
$14.0<r<21.7$ & $-2<r-i<2$ \\ 
$14.0<i<20.8$ & $-2<i-z<2$ \\ 
$14.0<z<20.0$ & 
\end{tabular}

It needs to be stressed that the highly dishomogeneous distribution of the
objects in the redshift space implies that the density of the training
points dramatically decreases for increasing redshifts, and that:\ i) unless
special care is paid to the construction of the training set, all networks
will tend to perform much better in the range where the density of the
training points is higher;\ ii) the application to the photometric data set
will be strongly contaminated by the spurious determinations.

\subsection{The construction of the training set}

In order to achieve an optimal traning of the NNs, two different approaches
to the construction of the training, validation and test sets were
implemented:\ the uniform sampling and the clustered sampling (via K-means
and/or SOM).

In both cases the training set data are first ordered by increasing
redshift, then, in the case of uniform sampling, after fixing the number of
training objects (which needs in any case to be smaller than 1/3 of the
total sample) objects are extracted following a decimation procedure. This
approach however, is undermined by the fact that the input parameter space
is not necessarily uniformously sampled, thus causing a loss in the
generalization capabilities of the network.

\begin{table*}
\textbf{Table 2.}{\small \ Column 1: higher accepted spectroscopic redshift
for objects in the training set;\ column 2:\ input (hence number of input
neurons)\ parameters used in the experiment;\ column 3:\ number of neurons
in the hidden layer;\ column 4:\ interquartile errors evaluated on the test
set;\ column 5:\ number of objects used in each of the training, validation
and test set.} \smallskip

\begin{tabular}{lllll}
\textbf{Range } & \textbf{parameters} & \textbf{neu.} & \textbf{error} & 
\textbf{objects} \\ 
{\footnotesize $z<0.3$} & {\footnotesize r, u-g, g-r, r-i, i-z} & 
{\footnotesize 18} & {\footnotesize 0.029} & {\footnotesize 12000} \\ 
{\footnotesize $z<0.5$} & {\footnotesize r, u-g, g-r, r-i, i-z} & 
{\footnotesize 18} & {\footnotesize 0.031} & {\footnotesize 12430} \\ 
{\footnotesize $z<0.7$} & {\footnotesize r, u-g, g-r, r-i, i-z} & 
{\footnotesize 18} & {\footnotesize 0.033} & {\footnotesize 12687} \\ 
&  & ${}$ &  &  \\ 
{\footnotesize $z<0.3$} & {\footnotesize r, u-g, g-r, r-i, i-z,
radius} & {\footnotesize 18} & {\footnotesize 0.025} & {\footnotesize 12022}
\\ 
{\footnotesize $z<0.5$} & {\footnotesize r, u-g, g-r, r-i, i-z,
radius} & {\footnotesize 18} & {\footnotesize 0.026} & {\footnotesize 12581}
\\ 
{\footnotesize $z<0.7$} & {\footnotesize r, u-g, g-r, r-i, i-z,
radius} & {\footnotesize 18} & {\footnotesize 0.031} & {\footnotesize 12689}
\\ 
&  &  &  &  \\ 
{\footnotesize $z<0.3$} & {\footnotesize r, u-g, g-r, r-i, i-z,
radius, petrosian fluxes, surface brightness} & {\footnotesize 22} & 
{\footnotesize 0.020} & {\footnotesize 12015} \\ 
{\footnotesize $z<0.5$} & {\footnotesize r, u-g, g-r, r-i, i-z,
radius, petrosian fluxes, surface brightness} & {\footnotesize 22} & 
{\footnotesize 0.022} & {\footnotesize 12536} \\ 
{\footnotesize $z<0.7$} & {\footnotesize r, u-g, g-r, r-i, i-z,
radius, petrosian fluxes, surface brightness} & {\footnotesize 22} & 
{\footnotesize 0.025} & {\footnotesize 12680}
\end{tabular}
\end{table*}

In the clustered sampling method, objects in each redshift bin are first
passed to a SOM\ or a K-means algorithm which performs an unsupervised
clustering in the parameter space looking for the most significant
statistical similarities in the data. Then, in each bin and for each
cluster, objects are extracted in order to have an uniform sampling of the
parameter space. This second procedure, while being slower than the uniform
sampling allows a complete and statistically homogeneous coverage of the
parameter space.

\subsection{The photometric redshift evaluation}

In order to evaluate the performances of the software as close as possible
to the detection limit of the data, we did not introduce strong cuts on the
limiting magnitudes. Hence the filters listed in Table 1, were applied to
the magnitudes and to the colors. The latter were introduced in order to
remove a few spurious objects present in the original data set.

The experiments were performed using the NNs in the Matlab and Netlab
Toolboxes, with and without the Bayesian framework. All NNs had only one
hidden layer and the experiments were performed varying the number of the
input parameters and of the hidden units. Extensive experiments lead us to
conclude that the Bayesian framework provides better generalization
capabilities with a lower risk of overfitting, and that an optimal
compromise between speed and accuracy is achieved with a maximum of 22
hidden neurons and 10 Bayesian cycles.

In Table 2, we summarize some of the results obtained from the experiments
and, in Figure 4, we compare the spectroscopic redshifts versus the
photometric redshifts derived for the test set objects in the best
experiment.

\subsection{Contamination of the catalogues}

In practical applications, one of the most important problems to solve is
the evaluation of the contamination of the final photometric redshift
catalogues or, in other words, the evaluation of the number of objects which
are erroneously attributed a $z_{phot}$ significantly (accordingly to some
arbitrarily defined treshold) different from the unknown $z_{spec}$. This
problem is usually approached by means of extensive simulations. The problem
of contamination is even more relevant in the case of NNs based methods,
since NNs are necessarily trained only in a limited range of redshifts and,
when applied to the real data, they will produce misleading results for most
(if not all)\ objects which ''in the real word'' have redshifts falling
outside the training range. This behaviour of the NNs is once more due to
the fact that while being good interpolation tools, they have very little,
if any, extrapolation capabilities. Furthermore, in mixed surveys, the
selection criteria for the spectroscopic sample tend to favour the brightest
(and, on average, the closer) galaxies with respect to the fainter and more
distant ones and, therefore, the amount of contamination encountered, for
instance, in the test set sets only a lower limit to the percentage of
spurious redshifts in the final catalogue.

\begin{figure*}
   \centering
   \includegraphics[width=10cm]{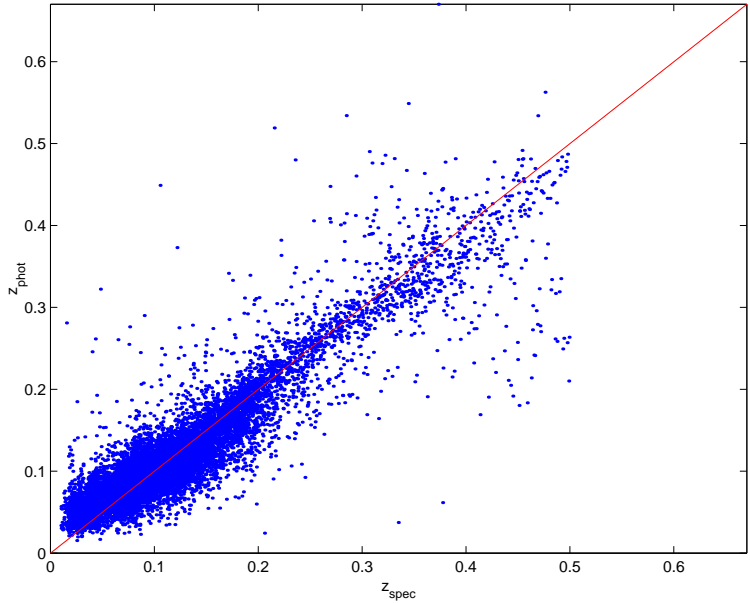}
     \caption{Photometric versus spectroscopic redshifts obtained with a Bayesian 
MLP with 2 optimization cycles, 50 learning epochs of quasi-Newton algorithm and 5 
inner epochs for hyperparameter optimization. Hyperparameters were initialized 
at $\protect% 
\alpha $=0.001 and $\protect\beta $=50}
     %\label{FigGam}%
\end{figure*}

To be more specific:\ in the SDSS-EDR spectroscopic sample, over a total of
54,008 objects having $z>0$, only $88\%$, $91\%$ and $93\%$\ have redshift $%
z $ lower than, respectively than $0.3$, $0.5$ and $0.7$. To train the
network on objects falling in the above ranges implies , respectively, a
minimum fraction of $12\%$, $9\%$ and $7\%$ of objects in the photometric
data set having wrong estimates of the photometric redshift. On the other
hand, as we have shown, the higher is the cut in redshifts, the lower is the
accuracy and a compromise between these two factors needs to be found on
objective grounds.

\begin{figure*}
   \centering
   \includegraphics[width=14cm]{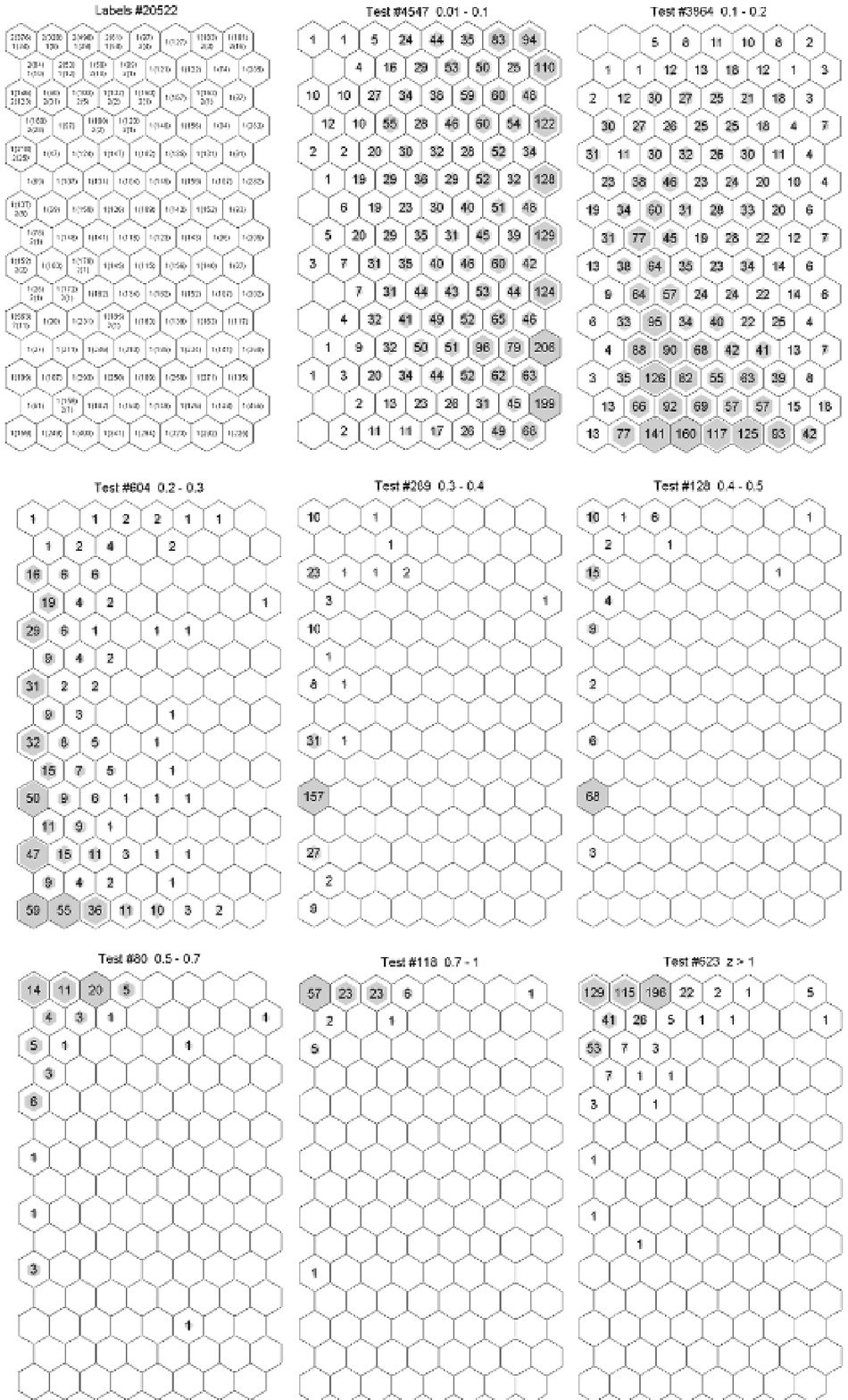}
     \caption{Maps of the neuron
activated by the input data set. As before, exagons represent the NN\ nodes.
In the map in the upper left corner, for a given node, the figures n(m)\ can
be read as follows: n is the class (n=1 meaning $z<0.5$ and n=2
meaning $z>0.5$) and m is the number of input vector of the
correspondent class which have activated that node. This map was produced
using the training and validation data sets. The other maps, produced each
in a different redshift bin, indicate how many input vector from the test
data set activated a given node.}
     %\label{FigGam}%
\end{figure*}

An accurate estimate of the contamination may be obtained using unsupervised
SOM clustering techniques over the training set.

In Figure 5 we show the position of the BMU as a function of the
redshift bin. Each exagon represents a neuron and it is clearly visible that
low redshift ($z<0.5$)\ tend to activate neurons in the lower right part of
the map, intermediate redshift ones ($0.5<z<0.7$) neurons in the lower left
part and, finally, objects with redshift higher than $0.7$ activate only the
neurons in the upper left corner. The labeling of the neurons (shown in the
upper left map) was done using the training and validation data sets in
order to avoid overfitting, while the confidence regions were evaluated on
the test set.

Therefore, test set data may be used to map the neurons in the equivalent of
confidence regions and to evaluate the degree of contamination to be
expected in any given redshift bin. Conversely, when the network is applied
to real data, the same confidence regions may be used to evaluate whether a
photometric redshift correspondent to a given input vector may be trusted
upon or not.

The above derived topology of the network is also crucial since it allows to
derive the amount of contamination. In order to understand how this may be
achieved, let us take the NN\ whose topological properties are shown in
Figure 5, and consider the case of objects which are attributed a redshifts $%
z_{phot}$ $<0.5$. This prediction has a high degree of reliability only if
the input vector activates a node in the central or right portions of the
map. Vector producing a redshift $z_{phot}$ $<0.5$ but activating a node
falling in the upper left corner of the map are likely to be misclassified.
In Figure 6, we plot the photometric versus spectroscopic redhift for all
test set objects having $z_{phot}$ $<0.5$ and activating nodes in the
correct region of the map.

As it can be seen, out of 9270 objects with $z_{phot}$ $<0.5$, only 39 (%
\textit{id est}, 0.4\% of the sample) have discordant spectroscopic
redshift. A confusion matrix helps in better quantifying the quality of the
results. In Table 3, we give the confusion (or, in this case,
'contamination') matrix obtained dividing the data in three classes
accordingly to their spectroscopic redshifts, namely class I:\ $0<z<0.3$,
class II:\ $0.3<z<0.5$, class III:\ $z>0.5$. The elements on the diagonal
are the correct classification rates, while the other elements give the
fraction of objects belonging to a given class which have been erroneously
classified into an other class.

\begin{table}
\textbf{Table 3}:\ confusion matrix for the three classes described in the text.

\begin{tabular}{lllll}
& objects & Class I & Class II & Class III \\ 
Class I & $9017$ & $95.4\%$ & $2.96\%$ & $1.6\%$ \\ 
Class II & $419$ & $6.4\%$ & $76.6\%$ & $16.9\%$ \\ 
Class III & $823$ & $3.8\%$ & $2.1\%$ & $94.2\%$%
\end{tabular}
\newline
\end{table}

As it can be seen, in the redshift range $(0,0.3)$, 95.4\% of the objects
are correctly identified and only $4.6\%$ is attributed a wrong redshift
estimate. In total, $94.2\%$ are correctly classified. By taking into
account only the redshift range $0<z<0.5$, this percentage becomes $97.3\%$.
From the confusion matrix, we can therefore derive a completeness of $97.8\%$
and a contamination of about $0.5\%$.

A simple step filter applied to the accepted BMU's allows therefore to
optimise the filter performances.\ For instance, it allows to choose whether
to minimize the number of misclassified objects (thus reducing the
completeness) or to minimize the classification error in a given redshift
bin more than in another one.

Another possible use of the topological properties of the SOM\ will be
discussed in a forthcoming paper and concerns the use of BMU\ to choose for
each given input vector the optimal NN.

\section{Summary and conclusions}

The application of NNs to mixed data, \textit{id est} spectroscopic and
photometric surveys, allows to derive photometric redshifts over a wide
range of redshifts with an accuracy equal if not better to that of more
traditional techniques.

The method makes use of three different neural tools:\ i)\ an unsupervised
SOM used to cluster the data in the training, validation and test set in
order to ensure a complete coverage of the input parameter space;\ ii)\ a
MLP in Bayesian framework used to estimate the photometric redshifts; iii)\
a supervised SOM\ used to derive the completeness and the contamination of
the final catalogues. On the SDSS-EDR, the best result (interq. error =\
0.020) was obtained by a MLP with 1 hidden layer of 22 neurons, after 5
Bayesian cycles. Once they are trained, NNs are extremely effective in terms
of computational costs (the 16 million objects in the SDSS-EDR are processed
in less than 50 min on a laptop;\ Longo et al. 2002, in preparation).
\begin{figure}
   \centering
   \includegraphics[width=8cm]{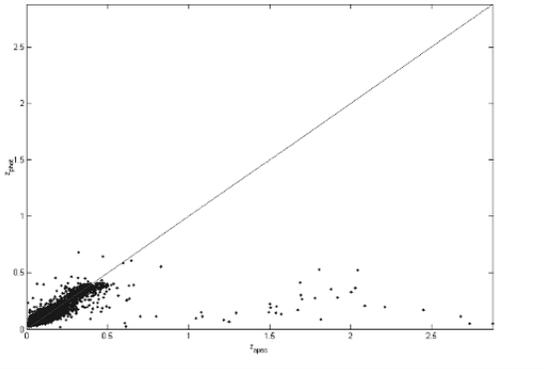}
     \caption{Spectroscopic versus photometric redshifts for objects belonging to 
Class I and Class II. Outliers are misclassified galaxies.}
     %\label{FigGam}%
\end{figure}

The method fully exploits the wealth of data provided by the new digital
surveys since it allows to take into account not only the fluxes, but also
the morphological and photometric parameters.

The proposed method will be particularly effective in mixed surveys, id est,
in surveys were a large amount of multiband photometric data is complemented
by a small subset of objects having also spectroscopic redshifts. It needs
also to be stressed that the foreseen implementation of the Virtual
Observatory will provide an ideal framework to NN based data mining tools:\
the availability of continuously updated data sets from where to extract
reliable and extensive training sets will allow a widespread use of
supervised NNs for the fast and accurate derivation of secondary parameters
such as the photometric redshifts.


\bigskip

\begin{thebibliography}{Le F\`{e}vre et al. 2000}
\bibitem[Allende Prieto et al. 2000]{Allende Prieto et al. 2000}  Allende
Prieto C., Rebolo R., Lopez R.J.G., Serra-Ricart M., Beers T.C., Rossi S.,
Bonifacio P., Molaro P., 2000, AJ, 120, 1516

\bibitem[Andreon et al. 2001]{Andreon et al. 2001}  Andreon S., Gargiulo G.,
Longo G., Tagliaferri R., Capuano N., 2000, MNRAS, 319, 700

\bibitem[Bailer and Jones 1998]{Bailer and Jones 1998}  Bailer-Jones C.A.L.,
Irwin M., von Hippel T., 1998, MNRAS, 298, 361

\bibitem[Baum 1962]{Baum 1962}  Baum W.A., 1962, Problems of extragalactic
research, IAU\ Symp. n.15, 390

\bibitem{Bertin and Arnout 1996}  Bertin E., Arnout S., 1996, AAS, 117, 393

\bibitem[Bishop 1995]{Bishop 1995}  Bishop C.M., 1995, Neural Networks for
Pattern Recognition, Oxford University Press

\bibitem[Brunner et al. 2000]{Brunner et al. 2000}  Brunner R.J., Szalay
A.S., Connolly A.J., 2000, ApJ, 541, 527

\bibitem[Bruzual and Charlot 1993]{Bruzual and Charlot 1993}  Bruzual A.G.,
Charlot S., 1993, ApJ, 405, 538

\bibitem[Connolly et al. 1995]{Connolly et al. 1995}  Connolly A.J., Csabai
I., Szalay A.S., Koo D.C., Kron R.G., Munn J.A., 1995, AJ, 110, 2655

\bibitem[Connolly et al. 1998]{Connolly et al. 1998}  Connolly A.J., Szalay
A.S., Brunner R.J., 1998, ApJ, 499, L125

\bibitem[Fernandez-Soto et al. 2001]{Fernandez-Soto et al. 2001}  %
Fernandez-Soto A., Lanzetta K.A., Chen H.W., Pascarelle S.M., Yakate N.,
2001, ApJSS, 135, 41

\bibitem[Giordano 2001]{Giordano 2001}  Giordano G., 2001, '' A neural
technique for the analysis of cosmic large scale structure'', Laurea Thesis,
November 2001, University of Salerno

\bibitem[Kohonen 1995]{Kohonen 1995}  Kohonen T., 1995, Self-Organizing
Maps, Springer:Berlin-Heidelberg

\bibitem[Koo 1999]{Koo 1999}  Koo D.C., 1999, astro-ph/9907273

\bibitem[Lahav et al. 1996]{Lahav et al. 1996}  Lahav O., Naim A., Sodr\'{e}
L. jr., Storrie-Lombardi M.C., 1996, MNRAS, 283, 207

\bibitem[Le F\`{e}vre et al. 2000]{Le Fèvre et al. 2000}  Le F\`{e}vre et
al. 2000, in Clowes R.G. et al. eds., ASP. Conf. Ser., New era of Wide Field
Astronomy, 232, 449

\bibitem[Longo et al. 2001]{Longo et al. 2001}  Longo G., Tagliaferri R.,
Sessa S., Ortiz P, Capaccioli M., Ciaramella A., Donalek C., Raiconi G.,
Staiano A., Volpicelli A., in Astronomical Data Analysis, J.L. Stark and F.
Murtagh eds., SPIE n. 4447, p.61

\bibitem[MacKay et al. 1994]{MacKay et al. 1994}  MacKay et al., Bayesian
methods for backpropagation networks models of neural network III, 1994, New
York:Springer-Verlag

\bibitem[Massarotti et al. 2001a]{Massarotti et al. 2001a}  Massarotti M.,
Iovino A., Buzzoni A, 2001a, AA, 368, 74

\bibitem[Massarotti et al. 2001b]{Massarotti et al. 2001b}  Massarotti M.,
Iovino A., Buzzoni A., Valls-Gabaud D., 2001b, AA, 380, 425

\bibitem[Nabney and Bishop 1998]{Nabney and Bishop 1998}  Nabney I.T.,
Bishop C.M., 1998, Netlab:\ Neural Network Matlab Toolbox, Aston University

\bibitem[Neal 1994]{Neal 1994}  Neal, R.M., Bayesian Learning for Neural
Networks, Ph.D. thesis, 1994, University of Toronto (Canada)

\bibitem[Pushell et al. 1982]{Pushell et al. 1982}  Pushell J.J., owen F.N.,
Laing R.A., 1982, ApJ, 275, L57

\bibitem[Storrie-Lombardi et al. 1992)]{Storrie-Lombardi et al. 1992}  %
Storrie-Lombardi M.C., Lahav O., Sodr\'{e} L. jr, Storrie-Lombardi L.J.,
1992, MNRAS, 259, 8

\bibitem[Stoughton et al. 2001]{Stoughton et al. 2001}  Stoughton C., Lupton
R.H., Bernardi M., Blanton M. R., et al., 2001, AJ, 123, 485

\bibitem[Tagliaferri et al. 2002]{Tagliaferri et al. 2002}  Tagliaferri R.,
Longo G., Milano L., Ciaramella A., Donalek C., Raiconi G., Volpicelli A.,
2002, Neural Networks, in preparation

\bibitem[Vesanto 1997]{Vesanto  1997}  Vesanto J., 1997, Ph.D. Thesis,
Helsinky University of Technology

\bibitem[Wang et al. 1998]{Wang et al. 1998}  Wang Y., Bachall N., Turner
E.L., 1998, AJ, 116, 2081

\bibitem[Weaver 2000]{Weaver 2000}  Weaver W.B., 2000, ApJ, 541, 298

\bibitem[York et al. 2000]{York et al. 2000}  York D.G., et al. 2000, AJ,
120, 1579
\end{thebibliography}
\end{document}